\documentclass[12pt]{iopart}

\usepackage{graphicx}
\usepackage{dcolumn}
\usepackage{bm}
\usepackage{bbm}
\usepackage{epsfig}
\usepackage{mathrsfs}
\usepackage{color}
\newcommand{\bjdcomment}[1]{}
\def\COMMENTS{
\renewcommand{\bjdcomment}[1]{\\\textcolor{red}{\emph{##1}}\\}}
\COMMENTS
\newcommand{\eocomment}[1]{}
\def\COMMENTS{
\renewcommand{\eocomment}[1]{\\\textcolor{blue}{\emph{##1}}\\}}
\COMMENTS

\newcommand{\sub}[2]{{#1}_{ \mbox{\scriptsize #2}}}

\newcommand{\bv}[1]{\mathbf{ #1 }}

\def\beq{\begin{eqnarray}}
\def\eeq{\end{eqnarray}}

\def\CR{\nonumber\\[0.15cm]}
\newcommand{\rref}[1]{Ref.~\cite{#1}}

\begin{document}


\title[Formation of bright solitons in the collapse of a BEC]
{Dynamical formation and interaction of bright solitary waves and solitons in the collapse of 
Bose-Einstein condensates with attractive interactions}
\author{B. J. D\c{a}browska-W{\"u}ster$^{1}$\footnote{Present address: Centre for Theoretical Chemistry and Physics, Institute of Fundamental Sciences, Massey University Albany, North Shore MSC, 0745 Auckland, New Zealand}, S. W\"uster$^{2,3}$ and M.~J.~Davis$^{1}$}
\address{$^1$ The University of Queensland, School of Physical Sciences, ARC Centre of Excellence for Quantum-Atom Optics, Qld 4072, Australia}
\address{$^2$ The University of Queensland, School of Physical Sciences, Qld 4072, Australia}
\address{$^3$ Max Planck Institute for the Physics of Complex Systems, N\"othnitzer Strasse 38, 01187 Dresden, Germany}
\ead{b.dabrowska-wuester@massey.ac.nz}



\begin{abstract}
We model the dynamics of formation of multiple, long-lived, bright  solitary waves in the collapse of Bose-Einstein condensates with attractive interactions as studied
in the experiment of Cornish~\emph{et al.}~[Phys.~Rev.~Lett.~{\bf 96} 170401 (2006)]. Using both mean-field and quantum field simulation techniques, we 
find that while a number of separated wave packets form as observed in the experiment, they do not have a repulsive $\pi$ phase difference that has been previously inferred. We observe that the inclusion of quantum fluctuations causes soliton dynamics to be predominantly repulsive in one dimensional simulations independent of their initial relative phase.  However, indicative three-dimensional simulations do not support this conclusion and in fact show that quantum noise has a negative impact on bright solitary wave lifetimes.
Finally, we show that condensate oscillations, after the collapse, may serve to deduce three-body recombination rates, and that the remnant atom number may still exceed the critical number for collapse for as long as three seconds independent of the relative phases of the bright solitary waves.
\end{abstract}

			     
%
\section[Introduction]{Introduction\label{introduction}}

At low temperatures it is well known that the wave function of an interacting  Bose-Einstein condensate can be described by solutions of the nonlinear Gross-Pitaevskii equation (GPE) for both repulsive and attractive atomic interactions \cite{stringari:review}.  In a one dimensional system a Bose-Einstein condensate (BEC) with intrinsic attractive interactions can form localised states for any number of particles.  These are known as bright matter-wave solitons, and demonstrate particle-like behaviour in collision events~\cite{book:solitons}.   A bright matter-wave soliton has been observed in a quasi-1D geometry at ENS through the tuning of atomic interactions from repulsive to attractive in a $^7$Li condensate with a Feshbach resonance \cite{khay:brighsol}.  A bright soliton was also observed by Eiermann \emph{et al.}\ by tuning the effective mass of a repulsive $^{87}$Rb condensate to be negative with a 1D optical lattice~\cite{gap_exp}.

Corresponding soliton solutions for condensates with attractive interactions are not possible in free space in three dimensions.  However for a  harmonically trapped BEC a condensate wave function can be found for particle numbers less than a critical value $\sub{N}{cr}$ \cite{ruprecht:attractive}.  Condensates with a particle number exceeding this value are unstable against collapse as demonstrated in the Bosenova experiment at JILA~\cite{jila:nova} and in the observation of the formation of quasi-1D soliton trains at Rice University~\cite{strecker:brighsol}. Bright solitary wave (BSW) solutions  to the 3D GPE with moderately tight harmonic trapping in two dimensions can be found for a condensate with attractive interactions as demonstrated by Parker \emph{et al.}~\cite{parker_previouspreprint,parker_bsw1}.  These are analogous to the soliton solutions of the 1D GPE, but are only self-trapped in one dimension.  The formation and collisions of apparent multiple BSWs in condensate collapse have been observed recently by Cornish \emph{et al.}~\cite{jila:solitons}.  

In the Rice experiment, repulsive interactions between neighboring solitons have been inferred directly from the soliton trajectories~\cite{strecker:brighsol}.  In the JILA experiment, repulsive interactions between the BSWs were concluded indirectly as a consequence of the condensate atom numbers remaining in excess of $\sub{N}{cr}$~\cite{jila:solitons}.  The long-time evolution of soliton trains~\cite{li_rev,stoof_solitons, li_theory}
can qualitatively be accounted for by assuming that neighbouring solitons have a \emph{repulsive} phase difference $\Delta \varphi$. In mean-field theory this implies 
$\pi/2<\Delta \varphi<3\pi/2$~\cite{gordon_forces}.   However, the \emph{development mechanism} of a repulsive phase-relation between the soliton-like structures in the experiments has not been explained.

Here we attempt to understand {the physics of} how 3D bright solitary waves form in condensate collapse, by using mean-field and quantum dynamical simulation methods incorporating the effects of three-body loss. {We focus on the results of the experiment by}
Cornish \emph{et al.}~\cite{jila:solitons}, who caused an initially stable $^{85}$Rb condensate to collapse by switching the inter-atomic 
interaction from repulsive to attractive using a Feshbach resonance~\cite{mahir:review} whilst keeping the trapping potential on.
At this point the number of atoms in the condensate exceeds the critical atom number for stability, $\sub{N}{cr}$, which causes the macroscopic wave function to collapse~\cite{jila:nova,savage:coll,wuester:nova,wuester:nova2}, in the process losing atoms due to three-body recombination. 
{As was found in earlier experiments,} the onset of loss is sudden, and this allows the precise definition of the collapse time, $\sub{t}{collapse}$~\cite{jila:nova}, which is largely independent of the three-body recombination rate~\cite{jila:nova,savage:coll,wuester:nova}. Following this primary collapse, it might be expected that repeated collapse processes and subsequent loss occur until the atom number decreases below the critical value, $\sub{N}{cr}$~\cite{saito_ueda:intermitt}. 
In contrast,  experiments~\cite{jila:nova,jila:solitons} have reported remnant atom numbers, $\sub{N}{remn}$, 
several times larger than $\sub{N}{cr}$ even after long evolution times. In \rref{jila:solitons} this is attributed to the creation 
of multiple bright solitons with mutual repulsive phase differences preventing any further collapse of the condensate. 
Parker \emph{et al.}\ have studied the collisions of 3D bright solitary waves  for the conditions of the experiment of Cornish \emph{et al.}~\cite{jila:solitons} in mean-field theory as a function of relative phase and velocity \cite{parker_previouspreprint,parker_bsw} .  They have found that while the survival of the BSWs requires a repulsive relative phase at low velocities, for sufficiently large relative velocity the BSWs survive single collision events independent of the relative phase.  They also found that  a $\Delta \varphi$ in the repulsive range can inhibit collapse induced destruction of the soliton pair for many oscillation periods involving more than 40 collisions~\cite{parker_previouspreprint,parker_bsw}. 

{In this article we models the initial collapse and structure formation of the experiment of Cornish \emph{et al.}~\cite{jila:solitons} by solving the mean-field Gross-Pitaevskii equation in three dimensions with realistic parameters. }
We find that the dynamical \emph{creation} of solitons as in~\rref{jila:solitons} does \emph{not} favour repulsive $\Delta \varphi$. 
For the Rice experiment~\cite{strecker:brighsol} the absence of fixed repulsive phase-relations in Gross-Pitaevskii (GP) simulations of soliton creation was pointed out in~\rref{brand_solitons}. For the JILA experiment~\cite{jila:solitons}, the fact that the initial GP wave function has even symmetry prevents repulsive 
phase relations in the final state for an even number of BSWs as mean-field theory preserves this symmetry. In this situation the central two members of the train must therefore have $\Delta \varphi=0$. In order for this statement to be false, there must be some form of symmetry-breaking mechanism. 

\begin{figure}
\centering
\epsfig{file={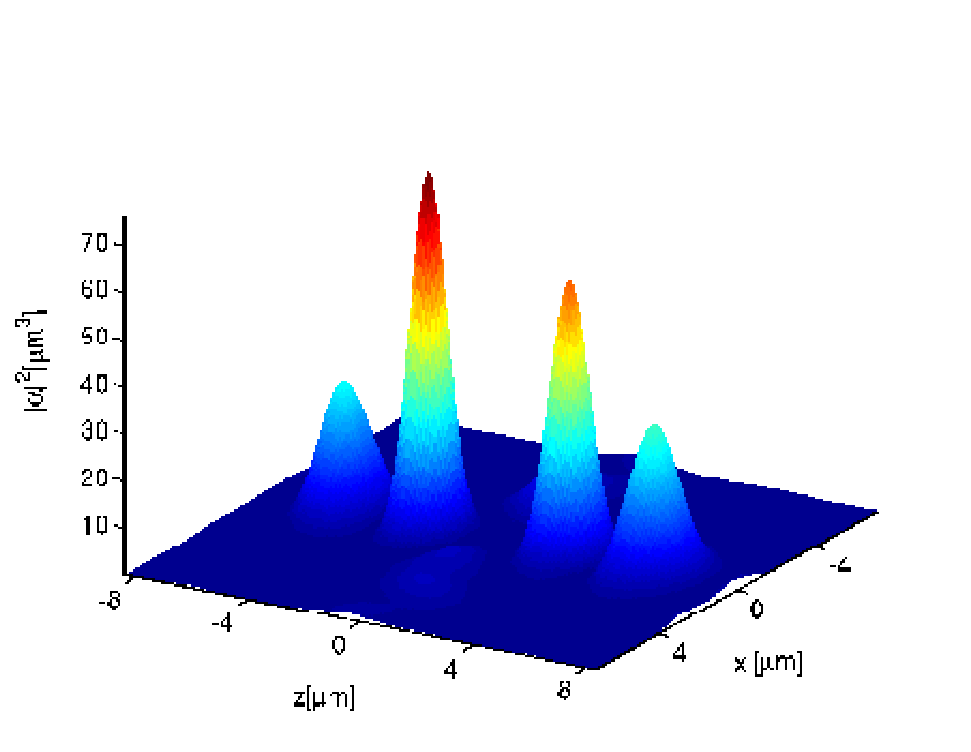},width=0.5\columnwidth} 
\caption{(colour online) Dynamically formed three-dimensional (3D) matter-wave BSWs (close-up of atom density cross-section for $y=0$).
Simulations using truncated Wigner quantum theory for $\sub{a}{collapse}=-20a_{0}$ and $K_{3}=5\times10^{-38}$~m$^{6}$s$^{-1}$. The snapshot is taken $16$~ms after the collapse. 
\label{solitons_figure}}
\end{figure}

To go beyond mean-field theory, we account for quantum field corrections to the GPE by modelling the collapse using the truncated 
Wigner approximation (TWA)~\cite{steel:wigner,Sinatra2001,castin:validity} which incorporates the effects of quantum noise.   However, we find that for conditions close to the experiment~\cite{jila:solitons} 
the relative phase between the central two dynamically formed BSWs is still near zero. An example of BSWs with attractive phase relations found within truncated Wigner quantum field theory is shown in \fref{solitons_figure}.

In order to better understand the effects of quantum noise, we simulate a 1D analogue of the JILA experiment~\cite{jila:solitons}, and find that quantum fluctuations have a significant effect on the resulting soliton interactions.  We perform controlled collisions between solitons of varying relative phase in 1D, and find that quantum noise tends to render soliton interactions more repulsive than the corresponding mean-field solution, \emph{irrespective} of their relative phase. 
 Three-dimensional (3D) simulations indicate that $\sub{N}{remn}$ can remain above $\sub{N}{cr}$ 
for long times in the presence of quantum and thermal fluctuations even if the interactions of the corresponding mean-field BSWs are attractive.  However, we find the effects of quantum noise tend to destroy the BSW structures for all initial relative phases, in contradiction to the results of mean field simulations \cite{parker_previouspreprint,parker_bsw} and experiment \cite{jila:solitons}.


%
%


\section{Three dimensional BEC collapse}
\label{collapse3d}
We first model collapsing condensates using the time-dependent Gross-Pitaevskii equation  (GPE) 
\begin{eqnarray}
\!\!\!\!\!\!\!\!\!\!\!\!\!\!\!\!\!\!
i \hbar \frac{\partial}{\partial t}  \Phi= 
\left(
 -\frac{\hbar^2}{2m}\nabla^{2}
+\frac{1}{2} m  ( \omega_r^2 r^2 +\omega_z^2 z^2 )  + g | \Phi |^2 - i \frac{\hbar}{2} K_2  | \Phi |^2- i \frac{\hbar}{2} K_3  | \Phi |^4 
\right) \Phi,
\label{GP}
\end{eqnarray}
with phenomenological two- and three-body loss terms~\cite{saito_ueda:intermitt,savage:coll, kagan:k3term}.
Here the condensate wave function $\Phi(r,z)$ is written in cylindrical co-ordinates. We assume $^{85}$Rb atoms with mass $m=1.411\times10^{-25}$ kg, interaction strengh $g=4\pi\hbar^{2}\sub{a}{s}(t)/m$, where the scattering length $\sub{a}{s}(t)$ varies in time. We define $\sub{a}{s}(t=0)=\sub{a}{ini}\geq0$ and $\sub{a}{s}(t>0)=\sub{a}{collapse}<0$. 
Further, $\omega_{r}/2\pi=17.3$ Hz, $\omega_{z}/2\pi=6.8$~Hz~\cite{jila:solitons}. The three-body loss rate, $K_{3}$, has not been precisely determined for the conditions of~\rref{jila:solitons}. For this reason, and for numerical necessity,  we vary $K_{3}$ throughout this work. We take $K_{2}=1.87\times 10^{-20}$~m$^{3}$s$^{-1}$~\cite{jila:K3measurement}. 
Our initial condition is the solution to the time-independent Gross-Pitaevskii equation with $\sub{a}{s}=\sub{a}{ini}$ found via imaginary time evolution. To reduce the computational demands of the problem we use a spatial grid size that does not fully accommodate burst atoms ejected during collapse~\cite{jila:nova}. As these are not the focus of our studies, we employ absorbing potentials near all numerical grid edges\footnote{The absorbers have the generic form $\sub{V}{damp}=i\gamma(1-\cos[\pi(d-\Delta x)/d])\theta(d-\Delta x)$, where $\Delta x$ is the distance to the nearest grid edge, $d$ the absorber width, $\theta$ the Heavyside step function and $\gamma$ is the absorber strength.}.

%
\begin{figure}
\centering
\epsfig{file={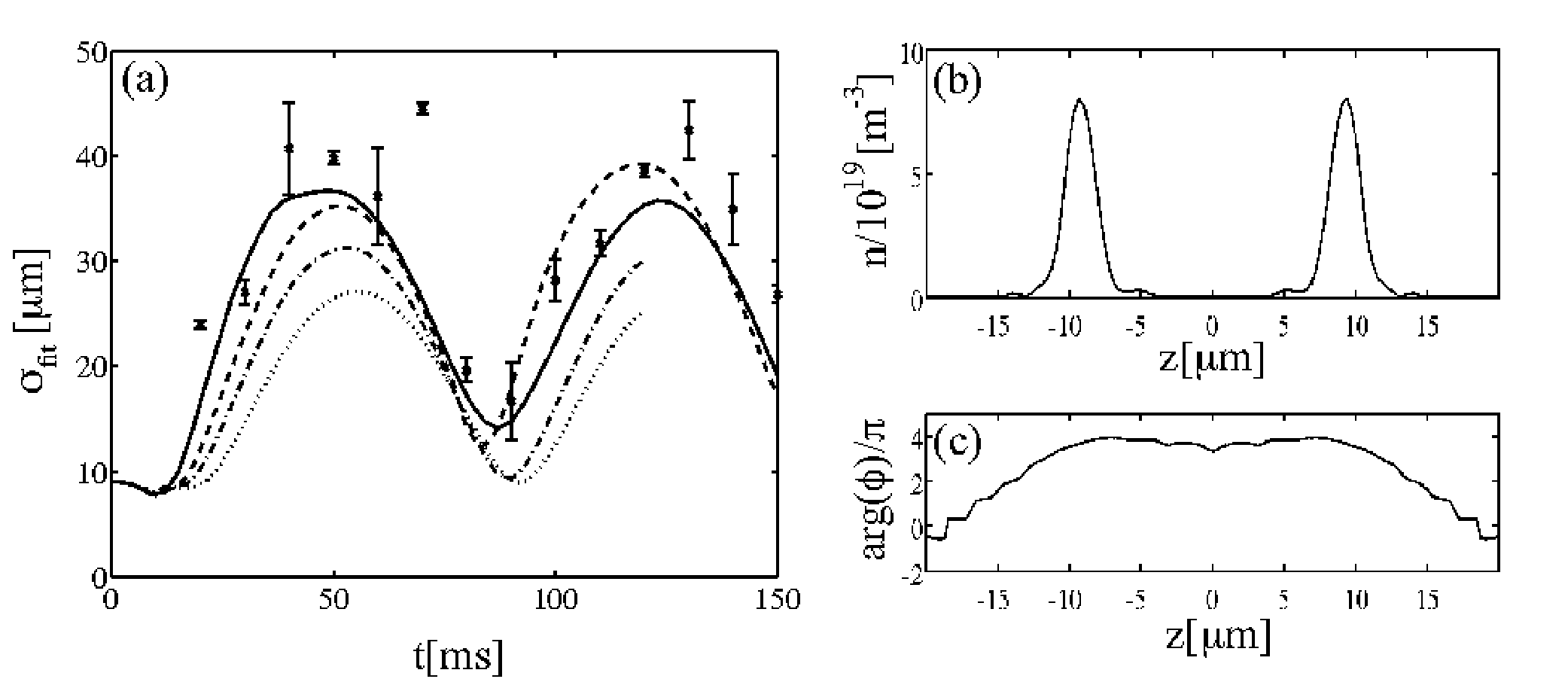},width=\columnwidth} 
\caption{Three-dimensional mean-field simulations of dynamically formed BSWs compared with experimental data from~\rref{jila:solitons}. (a) Full-width half-maximum (FWHM) in $z$ direction calculated for $a_{s}=-11.4 a_0$ and $K_{3}=0.2\times10^{-39}$~m$^{6}$s$^{-1}$ (solid line);  
$K_{3}=1\times10^{-39}$~m$^{6}$s$^{-1}$ (dashed); $K_{3}=4\times10^{-39}$~m$^{6}$s$^{-1}$ (dotted-dashed); $K_{3}=16\times10^{-39}$~m$^{6}$s$^{-1}$ (dotted).
Experimental points as in figure~3~of~\rref{jila:solitons} are shown with ($\times$).
For all results a 2D Gaussian function is fitted to the experimental and numerical column densities using a least squares method.
(b) Density and (c) phase of the condensate along the $z$-axis at the time $t=75$ ms.  
\label{fwhm_figure}}
\end{figure}

In~\fref{fwhm_figure} we compare our numerical solutions of \eref{GP} with the experimental data from~\rref{jila:solitons} for the case $\sub{a}{collapse}=-11.4 a_0$ and $\sub{a}{ini}=9 a_0$, where $a_{0}$ is the Bohr radius. For these initial conditions the experiment observes the formation of two BSW-like structures. We solve \eref{GP} in cylindrical co-ordinates using a Fourier transform based adaptive step-size Runge-Kutta algorithm.
The collapse time~\cite{jila:nova} for this scenario is $\sub{t}{collapse}=12$~ms. Near this time the peak-density rises dramatically with a maximal value that scales as $U/K_{3}$ \cite{saito_ueda:picture}. Directly after the collapse burst atoms are ejected from the condensate\footnote{In our simulations most of these are eliminated with the absorbing potential.}.
A so-called burst focus~\cite{jila:nova} takes place at a later time, when all ejected atoms that did not hit the absorber have classically completed half of an oscillation in the radial potential and have returned to the trap symmetry axis. This happens at $\sub{t}{bf}=\sub{t}{collapse} + \pi/\omega_{r}\approx41$~ms. 
We see the onset of the burst focus around $35$~ms. The burst focus develops into two axially separate clouds, in agreement with experimental observations~\cite{cornish:privcomm}. In our simulations these become precursors of a BSW pair.
It takes until $t=72$~ms before these clouds develop two distinct BSW-like shapes. 
The distance by which the atoms are axially displaced by the collapse, and hence the width found in~\fref{fwhm_figure}, increases for lower $K_3$ due to the larger interaction energy that is liberated during the collapse. Beyond $t\sim50$~ms the absorbing potentials become relevant for the dynamics. 

It is generally expected that the atom bursts have a strong uncondensed component~\cite{holland:burst, shlyapnikov:coll}, which in principle necessitates a theoretical treatment beyond the mean-field for $t>\sub{t}{collapse}$. The numerical results in~\fref{fwhm_figure} obtained using GP theory are hence approximate. 
Nonetheless, they describe the initial evolution of the experimentally observed full-width half-maximum (FWHM) rather well. Furthermore, the amplitude of the condensate oscillations depends strongly on the three-body loss rate~$K_{3}$ and the agreement between theory and experiment is better for smaller $K_{3}$. We suggest that this may allow for a better experimental measurement of $K_3$ than has been possible so far~\cite{jila:K3measurement}. The reliability of this approach would depend on the influence of the uncondensed burst atoms. 
Most importantly, we confirm that the BSWs are created \emph{in-phase} as required by symmetry. 

We now consider the effect of quantum fluctuations on the collapse. It has been suggested that quantum fluctuations can imprint a staggered phase pattern onto the condensate cloud during the development of instability~\cite{stoof_solitons}. To investigate this we employ the truncated Wigner approximation~\cite{wuester:nova2,steel:wigner,Sinatra2001,castin:validity,Blakie2005} 
of the full quantum evolution of the condensate. In this approach quantum field effects enter through the rigorous addition of noise to the initial condensate wave function
\begin{eqnarray}
\alpha(\bv{x},t=0)=\Phi(\bv{x},t=0) + \frac{1}{\sqrt{2}}\sum_{n}\varphi_{n}(\bv{x})\eta_{n}.
\label{TWAinistate}
\end{eqnarray}
Here $\Phi(\bv{x},t=0)$ is the condensate initial state as used in mean field simulations. $\alpha(\bv{x})$ is termed stochastic field, $\varphi_{n}(\bv{x})$ are the basis elements employed numerically and $\eta_{n}$ is complex Gaussian random noise with correlations $\overline{\eta_{n}\eta_{m}}=0$, $\overline{\eta^{*}_{n}\eta_{m}}=\delta_{nm}$.
In essence, this noise mimics the effect of vacuum fluctuations. The \emph{stochastic field} $\alpha(\bv{x})$ then evolves according to~\eref{GP}. For a more accurate treatment of the three-body losses we optionally include dynamical noise terms in our simulations~\cite{ashton:loss}. The theoretical and numerical methods employed here are described in~\rref{wuester:nova2}. The full truncated Wigner equation of motion for the stochastic field is~\cite{ashton:loss}
\begin{eqnarray}
d \alpha(\mathbf{x})
&=-\frac{i}{\hbar}\left(-\frac{\hbar^2}{2 m} \nabla^2 + V(\mathbf{x}) + \sub{U}{0}| \alpha(\mathbf{x})|^2\right)\! \alpha(\mathbf{x})dt
\CR
&-\frac{K_3}{2}| \alpha(\mathbf{x})|^4 \alpha(\mathbf{x})dt +\sqrt{\frac{3 K_3}{2}}| \alpha(\mathbf{x})|^2 d\xi(\mathbf{x},t),
\label{sgpe}
\end{eqnarray}
with dynamical noise $d\xi(x,t)$ that has correlations $\overline{d\xi(\mathbf{x},t)d\xi(\mathbf{x}',t')}=0$, $\overline{d\xi(\mathbf{x},t)^{*}d\xi(\mathbf{x}',t')}=\delta(\mathbf{x}-\mathbf{x}')\delta(t-t')$. Here, the cylindrical symmetry is broken by the noise in the initial state \eref{TWAinistate}. 
For a consistent treatment of the noise the propagation algorithm uses the harmonic oscillator basis~\cite{Blakie2005, wuester:nova2}. For the problem to remain computationally tractable in this basis we increase $K_{3}$ compared to the values in~\fref{fwhm_figure} (see e.g.~\fref{solitons_figure}), thus sacrificing direct correspondence to the experiment. 
We also apply the quantum treatment to a low energy subset of the full mode space in order to justify the TWA. Within these constraints our simulations show no sign of a strong preference for repulsive $\Delta \varphi$ within a large range of $\sub{a}{collapse}$. 
Therefore, the results of our 3D simulations question the assertion that BSWs originating from a condensate collapse as in~\cite{jila:solitons} are  created with a repulsive phase relation between neighbours. 

\section{One dimensional collapse with quantum noise}
\begin{figure}
\centering
\epsfig{file={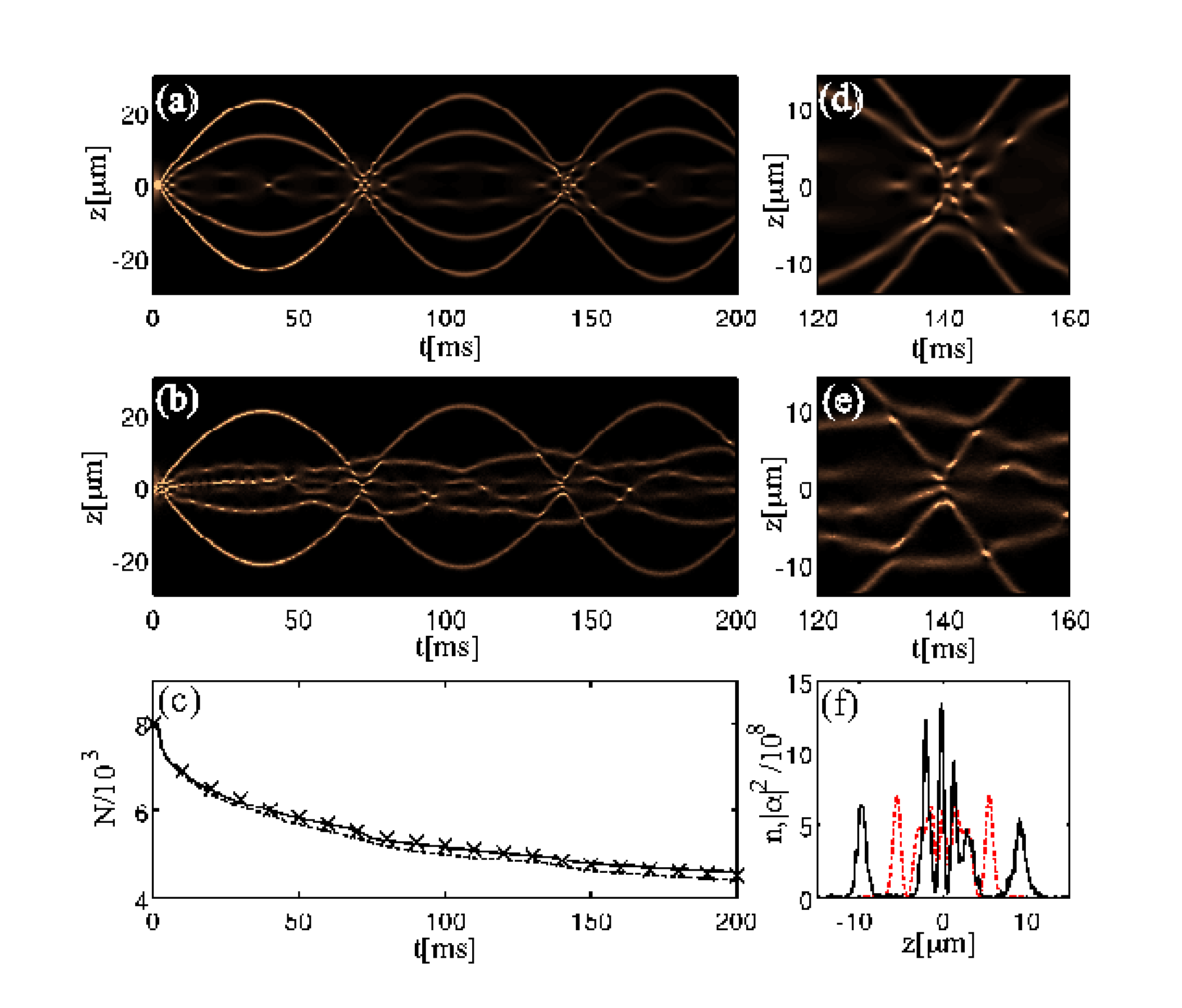},width=\columnwidth} 
\caption{(colour online) Time evolution of the atomic density (a-d,f) and atom number (e) in a one dimensional harmonic trap. We start from an initial total number of $\sub{N}{ini}=8000$ atoms in the trap ground state ($\sub{a}{ini}=0$), which is then exposed to an attractive nonlinearity with $\sub{a}{collapse}=-40 a_{0}$. A collapse occurs around $t=5$ ms. Other parameters are $\omega_{z}/2\pi=6.8$~Hz, $a_{r}=1$ $\mu$m ($\omega_r/2\pi=100$~Hz) and $K_{3}=2\times 10^{-40}$~m$^{6}$s$^{-1}$.
Bright (dark) regions correspond to high (low) atomic density. We truncate the density scale at $10\%$ of its peak value to improve visualisation. Shown are densities from (a,b) mean-field theory; (c,d) truncated Wigner approximation. Panels (b,d) show close-ups of the second overall contraction around $140$ ms. Panel~(f) shows the one dimensional GP density, $n$, (red-dashed) in comparison with the stochastic density, $|\alpha|^{2}$, (black-solid) at $t=140$ ms. 
The atom numbers in panel~(e) are from mean-field theory (solid line), TWA with initial noise only (dashed line) and TWA with dynamical noise ($\times$) with the noise contribution to the atom number subtracted for the truncated Wigner results.
\label{noise_repulsion}}
\end{figure}
While the investigation of quantum effects in three dimensions for realistic experimental parameters remains computationally intractable\footnote{The strong localisation of the condensate at the moment of collapse necessitates a fine numerical grid and the ejected burst atoms require a large spatial grid domain. Additionally, the experimental conclusions stem from BSW behaviour at long time. Altogether these requirements mean that quantum simulations for the precise experimental parameters would be extremely demanding.},
we can potentially gain insight into the effects of quantum noise using a one dimensional model. 
To do so we integrate out the transverse dimension $r$ of~\eref{GP} for $K_{2}=0$, using the approximation that it remains in the ground harmonic oscillator state \cite{bao:gpedimred}.
The 1D equation becomes
\begin{eqnarray}
i \hbar \frac{\partial}{\partial t}  \Phi=\left[-
\frac{\hbar^{2}}{2m}\frac{\partial^{2}}{\partial z^{2}} +  \frac{1}{2}m\omega_{z}^{2} z^{2} + \sub{g}{1D}|\Phi|^{2} -i\frac{\hbar}{2}\tilde{K}_{\mbox{\scriptsize 3,1D}}|\Phi|^{4} 
\right]\Phi.
\label{gp1d}
\end{eqnarray}
where $\sub{g}{1D}=g/(2\pi a_{r}^{2})$ and $\sub{\tilde{K}}{3,1D}=K_{3}/(3\pi^{2} a_{r}^{4})$ with $a_{r}=\sqrt{\hbar/(m\omega_{r})}$.

\Fref{noise_repulsion} shows typical simulations of a one dimensional collapse\footnote{We refer to a one dimensional collapse as the situation where a narrow density spike has developed and then exploded due to the onset of strong losses. Unlike in 3D, a 1D setting does not develop a mathematical density singularity.}. \Fref{noise_repulsion}(a) is the mean-field GPE solution.
Close inspection of the high density regions reveals soliton-like structures for most times. Their number, however, does not remain conserved at all times but varies after subsequent collapses. We find that no fixed repulsive phase-relations form and that the overall system exhibits breathing oscillations with repeated increase of density accompanied by three-body loss whenever solitons closely approach one another.

Applying the dimensionality reduction used above to \eref{sgpe} yields the one dimensional stochastic differential equation for the truncated Wigner framework
\begin{eqnarray}
 d\alpha&=&-\frac{i}{\hbar}\left[-
\frac{\hbar^{2}}{2m}\frac{\partial^{2}}{\partial z^{2}} +  \frac{1}{2}m\omega_{z}^{2} z^{2} + \sub{g}{1D}| \alpha |^{2}\right]\alpha  dt
\CR
&&-\frac{\sub{\tilde{K}}{3,1D}}{2}|\alpha|^{4} \alpha dt+  \sqrt{\frac{9 \tilde{K}_{3,1D}}{8}}| \alpha|^2 d\xi(x,t).
\label{sde1d}
\end{eqnarray}
\Fref{noise_repulsion}(c) shows a trajectory of a 1D collapse simulation using the TWA with initial and dynamical noise, implemented only on the central half of Fourier space to avoid aliasing and keep the noise density much lower than the condensate density~\cite{norrie:long}. A qualitative difference in the dynamics in the presence of vacuum fluctuations [\fref{noise_repulsion}(c)] can be noted compared to the mean-field simulations: The overall dynamics of the solitons is more repulsive leading to a decrease in the overall atom loss due to lower densities at the moments of solitons closest approach. This behaviour is typical for all realisations of noise. In the following section we investigate the phenomenon of noise induced repulsion more closely.

\section{Quantum soliton collisions} 

While we do not find any evidence that quantum noise facilitates the creation of solitons (in 1D) or BSWs (in 3D) with repulsive phase relations during the BEC collapse, our 1D model suggests that the noise strongly affects the collisional dynamics of the observed solitons. 
Strong effects of incoherence on the interaction properties of solitons are for example known from nonlinear optics \cite{andersen_incoherent,stegeman_collisions}.

\begin{figure}
\centering
\epsfig{file={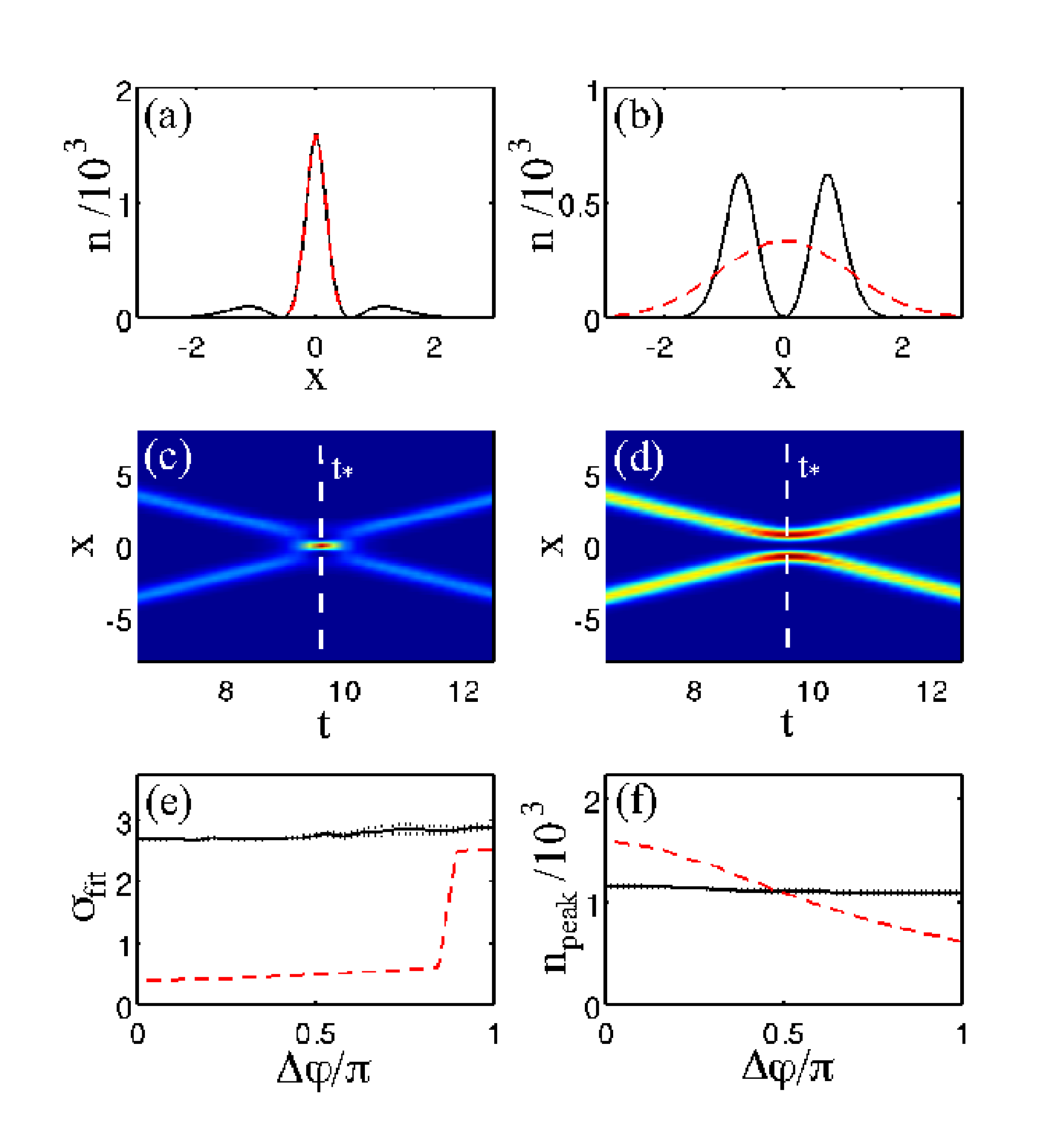},width=0.6\columnwidth} 
\caption{(colour online) Binary interactions of bright solitons in conservative mean-field and truncated Wigner models quantified by the full-width half-maximum (FWHM) in the longitudinal, $z$, direction.
Upper panels: Spatial soliton profiles (black-solid) and Gaussian fit (red-dashed) at time of closest approach, $t_{*}$, for $\Delta \varphi=0$ (a) and $\Delta \varphi=\pi$ (b). Middle panels:  Close-ups of time evolution of soliton densities (red highest, blue lowest) for $\Delta \varphi=0$ (c) and $\Delta \varphi=\pi$ (d). Lower panels: Width of Gaussian fit at $t_{*}$ vs. $\Delta \varphi$ (e)  including mean-field result (red-dashed) and TWA averaged over 200 realizations (black-solid). 
Overall peak density $\sub{n}{peak}$ is shown in panel (f) with lines as in (e). In (e,f) the sampling error of the stochastic averaging is indicated with dotted lines.
\label{1dcollisionplots}}
\end{figure}

To understand this in more detail, we focus on the collision stage and investigate the effect of quantum noise on the interactions of analytically prepared bright solitons. For this we make Eq.~\eref{gp1d} dimensionless with energy, time and length scales given by $\hbar\omega_{r}$, $\omega_{r}^{-1}$ and $a_{r}$ respectively, and set $\sub{\tilde{K}}{3,1D}=0$.
The initial state of these 1D simulations is a superposition of a pair
of separated bright solitons, individually multiplied by a phase factor $e^{ik_{n}x}$ such that the solitons  propagate towards one other with equal speed. Explicitly
\begin{equation}
\Phi(x,t=0)=A\sum_{n=1,2}\mbox{sech}{[\gamma(x-p_{n})]}e^{i(k_{n}x + \varphi_{n}) },
\end{equation}
with $A=20$, $\gamma=|\tilde{g}|A$, $p_{1,2}=\pm10$, $k_{1,2}=\mp1$ and $\varphi_1=0$.
Within mean-field theory the character of the ensuing collision is determined by the independent parameter of relative phase 
$\Delta \varphi=\varphi_{2}$. The density evolution for $\Delta \varphi=0$ ($\Delta \varphi=\pi$) is shown in~\fref{1dcollisionplots}(c) [\fref{1dcollisionplots}(d)]. 
The most significant difference between collisions with opposite relative phase occurs at the moment of closest approach of the solitons, which is labelled by $t_{*}$. The solitons merge into a single peak for $\Delta \varphi=0$, while for $\Delta \varphi=\pi$ they preserve a double peak structure.

\Fref{1dcollisionplots}(a,b) demonstrates a Gaussian fit to the overall density profile at the time of collision, giving a quantitative criterion to distinguish the attractive and repulsive situations.
In the attractive case the width of the Gaussian, $\sub{\sigma}{fit}$, is given by the single narrow solitonic peak [\fref{1dcollisionplots}(a)], while in the repulsive case it is set by the separation between the solitons, which is much larger [\fref{1dcollisionplots}(b)]. A second distinguishing feature is the peak density at the moment of closest approach, which is also the peak density during the collision $\sub{n}{peak}=\mbox{max}_{x,t}|\Phi(x,t)|^{2}$. The dependence of both these indicators on the relative phase $\Delta \varphi$ is shown in \fref{1dcollisionplots}(e,f). In the mean-field treatment they behave as expected, i.e.~there is a clear increase in $\sigma_{\rm fit}$ around $\Delta \phi = \pi$. However, the situation becomes completely different if the effects of vacuum fluctuations on these collisions are accounted for. 

We find that in the presence of noise the relative phase between the solitons in the mean-field component of the condensate wave function loses its significance. 
When averaged over 200 realisations of the atom-field with different realisations of noise, the overall peak of the density, $n_{\rm peak}$, and the width of the Gaussian fit, $\sub{\sigma}{fit}$, are no longer distinct for in-phase and out-of-phase solitons\footnote{In detail, 
we first obtain $\sub{\sigma}{fit}$ and $\sub{n}{peak}$ from each trajectory and \emph{then} take the average. Although, this does not have a direct interpretation within the TWA it provides an indication of how individual experimental realisations might behave and serves as a measure of the effect of quantum fluctuations on the soliton collisions.}.
Instead, both variables are \emph{independent} of the initial relative phase. For all relatives phases the width of the Gaussian fit corresponds closely to that of the \emph{repulsive} mean field simulations.

\section{Three dimensional bright solitary wave collisions} 

Finally, we investigate whether observations regarding soliton interactions in the one dimensional simulations carry over to three dimensional collisions of bright solitary waves.
As stated earlier, full simulations of the entire collapse experiment are numerically intractable. It is possible, however, to study binary collisions of three dimensional BSWs using the Gross-Pitaevskii equation~\cite{parker_previouspreprint,parker_bsw}. We extend the work of \rref{parker_previouspreprint,parker_bsw} by performing the same simulations incorporating quantum noise in the truncated Wigner approximation, as well as the effects of three-body loss. It has previously been shown in \rref{parker_bsw} that collisions of bright solitary waves in attractive BECs depend on the collision velocity. In particular, at low velocities a phase difference $\Delta \varphi=\pi$ can increase the possible number of repeated collisions before a significant degradation of the soliton shape. In this section we show that quantum and thermal noise have the opposite effect, reducing the number of collisions with preserved soliton shape.

Our initial binary soliton state is chosen to most closely fit our findings of \sref{collapse3d}. Thus, we employ $K_{3}=2\times 10^{-40}$ m$^{6}$s$^{-1}$, an initial total atom number $\sub{N}{ini}=3500$ and a soliton separation of about $d=32$~$\mu$m for $\sub{a}{collapse}=-11.4a_{0}$. We find the initial soliton state using a conjugate gradient method, while neglecting the axial trap. For the dynamical simulation to replicas of it are symmetrically placed around the origin on the $z$-axis with a relative phase $\Delta\varphi$. The axial trap is then switched on and the solitary waves are allowed to propagate towards the origin to collide~\cite{parker_previouspreprint,parker_bsw}.

\begin{figure}
\centering
\epsfig{file={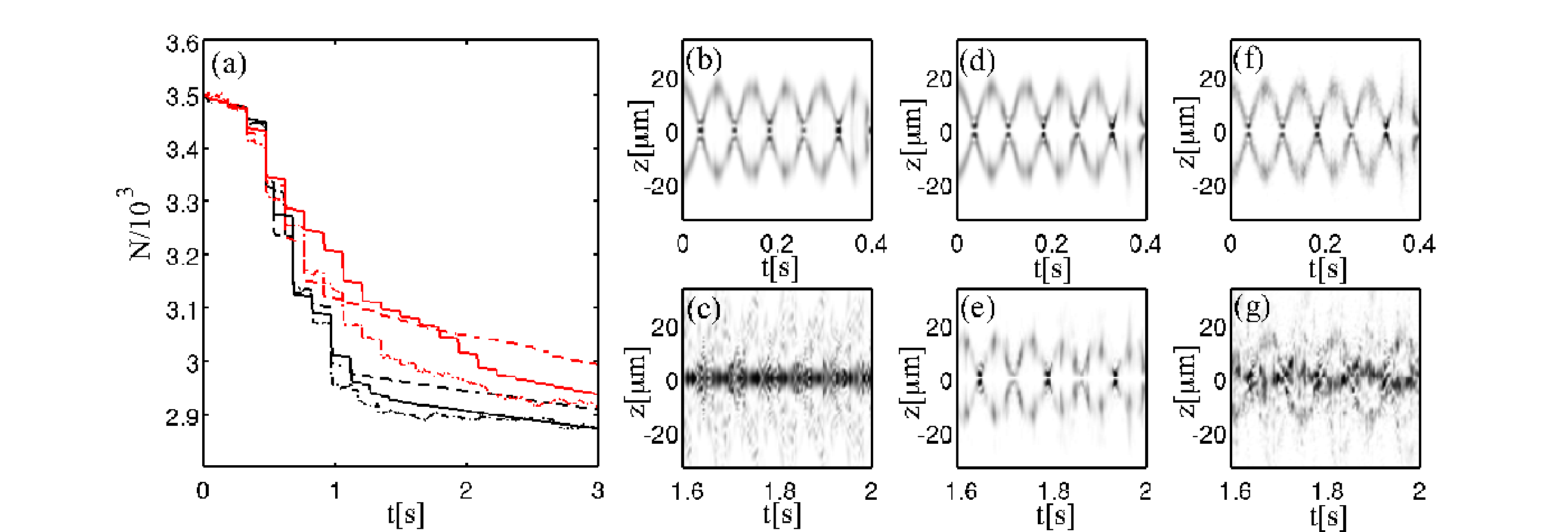},width=\columnwidth} 
\caption{(colour online) Binary collisions of three dimensional bright solitary waves under the influence of three-body loss (compare with Fig.~4 of Ref.~\cite{parker_previouspreprint} for corresponding data without three-body loss). (a) Total number of  atoms remaining in the trap: mean-field theory (solid), TWA with initial noise (dashed) and TWA with dynamical noise (dot-dashed). Results for different relative phases are shown: $\Delta\varphi=0$ (black, lower curves), $\Delta\varphi=\pi$ (red, upper curves). Panels (b-g) display atomic density along the $z$-axis with upper (lower) panels showing the evolution at early (late) times. We present mean-field simulations (b-e) and TWA with initial noise (f,g). The BSWs are in-phase $\Delta\varphi=0$ (b,c) or out of phase $\Delta\varphi=\pi$ (d-g). For BSWs within the single trajectory TWA we enforce the relative phase before quantum fluctuations are added. Darker areas correspond to higher densities. Note that for better visibility a nonlinear grey scale that exaggerates low density regions is used.
\label{3dcollisionplots}}
\end{figure}

We determine the remnant atom number of such a scenario after the evolution time of three seconds (as in the experiment of Cornish \emph{et al.}~\cite{jila:solitons}). Three theoretical approaches were employed for the dynamical evolution: mean-field theory, truncated Wigner formalism with initial noise only and TWA with the addition of dynamical noise~\cite{wuester:nova2}. All simulations use the harmonic oscillator basis.
The results are summarized in \fref{3dcollisionplots}.
It can be seen that 
for both initial phases the atom number remaining after $3$ seconds exceeds the critical number, which for this scenario is $\sub{N}{crit}\approx2800$.\footnote{The critical number is $\sub{N}{crit}=k \bar{a}/|a_{s}|$, with $k=0.55$, $\bar{a}=\sqrt{\hbar/m\bar{\omega}}$, $\bar{\omega}=(\omega_{r}^{2}\omega_{z})^{1/3}$ \cite{jila:solitons}.} 
Furthermore, in the presence of three-body loss [\fref{3dcollisionplots}(c,e)] the relative phase $\Delta\varphi=\pi$ allows oscillations of the BSW pair in the trap for longer times than $\Delta\varphi=0$, as was shown without the inclusion of three-body loss in \rref{parker_bsw}.

The inclusion of vacuum fluctuations via the truncated Wigner simulation does not significantly alter the remnant atom numbers around three seconds. It does, however, strongly affect the spatial BSW evolution. As shown in~\fref{3dcollisionplots}(g), even for $\Delta\varphi=\pi$ the two clearly separated clouds no longer survive at long evolution times, in contrast to mean-field studies [\fref{3dcollisionplots}(e)]. This also disagrees with the experimental results, which did observe BSW-like structures after three seconds and more than 40 collisions between the two wavepackets \cite{jila:solitons}.

Therefore, from our 3D simulations results we cannot conclude that the quantum noise renders BSW collisions  more repulsive --- in fact quantum noise appears to destabilise the case of $\Delta\varphi=\pi$.
We note, however, that this qualitative difference between 3D and 1D results could stem from the reduced amount of noise that we are forced to apply in 3D due to numerical reasons.\footnote{Care must be taken when interpreting these simulations as quantum field evolution in the sense of the TWA. Due to the application of noise to a restricted set of modes and, in particular, the long evolution times, the TWA is not strictly justified. None\-the\-less, the noise present in the simulation can closely mimic the effect of hot uncondensed atoms that are known to be present after the condensate collapse. In this sense we would expect the simulations with noise to offer a more realistic description than a pure GP treatment.} The  apparent agreement between mean field theory and the experimental results in \cite{parker_bsw} is compelling; however there is no obvious mechanism for the development of the $\pi$ phase difference, and we can offer no explanation for why a calculation that we expect to be more accurate than GP simulation gives results that so obviously differ from observations in the lab.

%
%

\section{Conclusions} 
We have studied the short and long time dynamics of the collapse of attractive Bose-Einstein condensates, applying 3D and effective 1D models that include three-body losses and quantum noise.
We find that due to the kinetic energy liberated, the condensate exhibits violent oscillations in the trap after the collapse, with an amplitude that strongly depends on three-body loss rates, $K_{3}$. We propose that the sensitivity of these oscillations after the collapse may improve measurements of $K_{3}$. After the collapse highly excited BSW-like structures form from the remnant condensate. These also exhibit violent dynamics and disappear, and reform multiple times throughout the evolution. The relative phase between dynamically formed BSWs does not usually
take repulsive values ($\pi/2<\Delta \varphi<3\pi/2$) as has been previously postulated~\cite{jila:solitons}. 
However, we show in one dimension that the addition of quantum fluctuations can render interactions repulsive, even for solitons with a phase relationship for which standard Gross-Pitaevskii theory predicts them to exhibit attractive interactions. Our three dimensional studies of interactions with quantum noise do \emph{not} show a similar effect at our level of approximation; instead they suggest that quantum noise results in a shorter BSW lifetime.
Our results raise questions about the interpretation of the JILA collapse experiment~\cite{jila:solitons} in terms of the creation of BSWs with mutually repulsive relative phases. In light of this we suggest that the system warrants further theoretical and experimental study. In particular, it will be interesting to compare the results of the present studies with
more direct treatments of quantum effects in soliton or BSW collisions. Experimentally, one could aim to measure the soliton phase directly \cite{weibin:phases} rather than inferring it from collisional behaviour.

\ack{We would like to thank Simon Cornish, Andy Martin, Nick Parker, and  Joachim Brand for helpful discussions and comments, and Simon Cornish for the provision of the experimental data.
This research was supported by the Australian Research Council Centre of Excellence for Quantum-Atom Optics and ARC Discovery Project DP0343094. We also thank the Queensland Cyber Infrastructure 
Foundation for providing computation resources at the National Facility of the Australian Partnership for Advanced Computing. BJDW is grateful for the hospitality of the Max-Planck 
Institute for the Physics of Complex Systems in Dresden.}


\end{document}